# The Insight-HXMT mission and its recent progresses

S. Zhang [a], S.N. Zhang [a,b], F.J. Lu [a], T.P. Li [a], L.M. Song [a], Y.P. Xu [a], H.Y. Wang [a], J.L. Qu [a], C.Z. Liu [a], Y. Chen [a], X.L. Cao [a], F. Zhang [a], S.L. Xiong [a], M.Y. Ge [a], Y.P. Chen [a], J.Y. Liao [a], J.Y. Nie [a], H.S. Zhao [a], S.M. Jia [a], X.B. Li [a], J. Guan [a], C.K. Li [a], J. Zhang [a], J. Jin [a], G.F. Wang [a], S.J. Zheng [a], X. Ma [a], L. Tao [a], Y. Huang [a]

[a] Key Laboratory of Particle Astrophysics, Institute of High Energy Physics, Beijing 100049, China
[b] University of Chinese Academy of Sciences, Beijing 100049, China

Further author information: (Send correspondence to S.Z.)
S.Z: E-mail: szhang@mail.ihep.ac.cn, Telephone: +86 10-88236114

## ABSTRACT

The Hard X-ray Modulation Telescope (HXMT or also dubbed as Insight-HXMT) is China's first astronomical satellite. It was launched on 15$^{th}$ June 2017 in JiuQuan, China and is currently in service smoothly. It was designed to perform pointing, scanning and gamma-ray burst (GRB) observations and, based on the Direct Demodulation Method (DDM), the image of the scanned sky region can be reconstructed. Here we introduce the mission and its progresses in aspects of payload, core sciences, ground calibration/facility, ground segment, data archive, software, in-orbit performance, calibration, background model, observations and preliminary results.

keywords: X-ray observations, calibration, background, data archive, software

## 1. INTRODUCTION

Since the launch of the first X-ray satellite Uhuru in 1970, the soft X-ray（below 20 keV) astronomy has developed explosively in the last century. Although at the hard X-rays (above 20 keV）energy, the first telescope HEAO-1 (A4) was launched in 1977, which was not much later than Uhuru, the corresponding developments in both detectors and astrophysics fell far behind that at soft X-ray energy. The main difficulty of the latter is lack of mature technology to collect hard X-ray photons via mirror reflection; therefore, hard X-ray observations are usually characterized with the high background and low sensitivity. Historically the development of hard X-ray telescopes went through mainly three stages. The first generation was in collimated mode, with typical effective area of at most 1000-2000 cm$^2$. To improve the imaging capability, hard X-ray telescopes was designed in coded-mask type since 2000, for which the angular resolution can reach arc-minutes level. Representatives are INTEGRAL and SWIFT/BAT, which have effective areas of roughly 3000-5000 cm$^2$ and large field of views (FOVs) of hundred to thousand deg$^2$. Finally, to improve further both the sensitivity and imaging capability, the first focusing hard X-ray telescope NuSTAR was launched in 2012, but with an effective area reduced to less than 200 cm$^2$ at above 20 keV.

The concept of Insight-HXMT was first proposed in 1994 and evolved into a mission covering 1-250 keV. The primary key science of Insight-HXMT was to carry out all-sky survey, for discovery of a large sample of the obscured supermassive black hole (BH), and to build an active galactic nuclei (AGN) catalogue which is essential for probing the extragalactic hard X-ray diffuse emission. This was actually a science window once only opened to the hard X-ray domain at the end of last century and the corresponding breakthroughs were already benefited from the launches of a series of hard X-ray missions, say, for example, the European INTEGRAL in 2002 and the American Swift in 2004. As a result, hard X-ray catalogues that contain thousands of sources were derived accordingly and smoothly along with the elapsed years since 2002. Obviously, although to build a hard X-ray catalogue of the extragalactic source was also the primary scientific target of Insight-HXMT when it was proposed by the end of last century, it did not hold anymore when Insight-HXMT was officially funded in 2011.

Fortunately, to build a powerful hard X-ray mission turns out to be never late given the lag of researches in hard X-rays to that in soft X-rays. One additional science window known all the time but so far less addressed is to observe simultaneously in broad band the X-ray binary systems (XRBs) in their bright states or outbursts. The difficulty lies in that it is usually hard to organize observational campaigns which cover simultaneously the soft X-rays and hard X-rays. In the meanwhile, the telescopes dedicated to pointing observations are usually suffered from pile-up problem in observing bright source at soft X-rays, and are short of effective area for collecting sufficient photons at hard X-rays. The contemporary missions before Insight-HXMT are all working at energies above a few keV and have relatively small effective area at hard X-rays. Apart from building for the first time a thorough sample for XRB outburst evolution at broad band, the advantage of having moderate FOV and large effective area of high energy telescope (HE) onboard Insight-HXMT will allow us to scan the Galactic plane and hunt for the short/faint transients which is another almost blank research field prior to Insight-HXMT.

As China's first astronomical satellite, Insight-HXMT was launched on 15th June 2017 in JiuQuan, China and is currently in service smoothly. After the launch, HXMT was officially dubbed as Insight-HXMT in honor of the Chinese famous physicist Prof. He ZeHui. Here we introduce in what follows the mission and its progresses, in aspects of payload, core sciences, ground calibration/facility, ground segment, data archive, software, in-orbit performance, calibration, background model, observations and preliminary results,.

## 2. CORE SCIENCES

The core sciences of the Insight-HXMT are:
(1) to scan the Galactic Plane and find new transient sources and to monitor the known variable sources;
(2) to observe X-ray binaries in broad band and study the dynamics and emission mechanism in strong gravitational or magnetic fields;
(3) to observe GRB at a relatively rarely explored band that covers a few hundred keV to a few MeV.

## 3. IMAGING TECHNIQUE

The DDM was proposed by Li and Wu in 1991[1], and has been successfully applied to analyzing the observational data of different types of telescopes [2,3]. The basic point of this algorithm is to solve the observational equations that are largely distorted by a variety of attributes, say for example the statistical errors. During the procedure of pursuing the proper solutions to this equation array, by introducing the physical constraints (for example the flux could not be negative) the distortions are largely suppressed and therefore one gets the desired solutions for both the flux and the position of the celestial object observed[1]. The Insight-HXMT Galactic survey map is produced with this method but by taking additionally an idea of properly handling/introducing background models as part of the response matrix in iteration procedure[4]: either the particle background or the known sources can be encoded into the transformed response matrix as background models, and tagged with scaling factors be estimated or fixed in the iteration procedure as 'pixels' additional to the inferred sky region.

## 4. PAYLOAD

Insight-HXMT carries three slat-collimated instruments (see Figure 1): HE, Medium Energy X-ray Telescope (ME), and Low Energy X-ray Telescope (LE). HE consists of 18 NaI/CsI phoswich modules (main detectors) with a total geometrical area of about 5000 $cm^2$ in 20-250 keV. ME takes 1728 Si-PIN pixels which cover an energy range of 5-30 keV and ends up with a total geometrical area of 952 $cm^2$. LE adopts the swept charge device (SCD) as its detectors, which is sensitive in 1-15 keV with a total geometrical area of 384 $cm^2$.

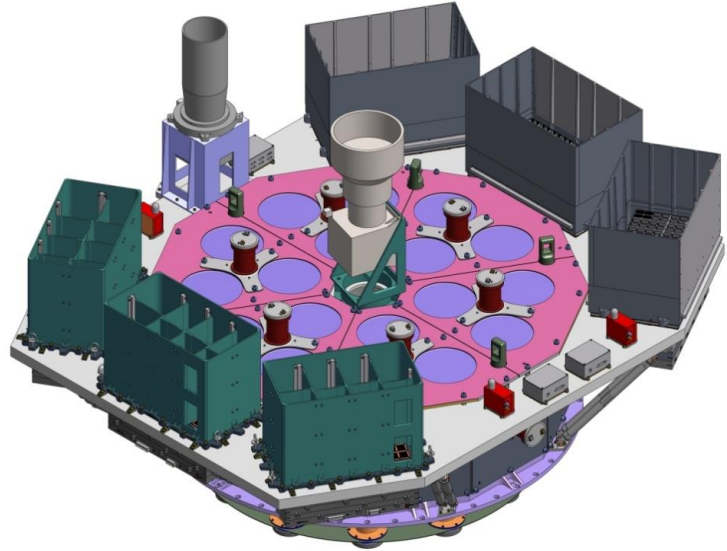

Figure 1: The main payloads onboard Insight-HXMT. The 18 cylindrical NaI/CsI detectors located at the center are HE, the boxes on the lower left are LE and upper right ME.

### 4.1 HE

With a cylindrical structure, HE encloses 18 phoswich modules and their collimators. The collection area of each detector is 283.5 cm$^2$, and its collimator defines FOVs of $1.1\times5.7$ deg$^2$ (for fifteen modules), $5.7\times5.7$ deg$^2$ for two modules, and one blind for measuring the background. The orientations of the 18 FOVs are formed in three groups, which change in a step of 60 deg.

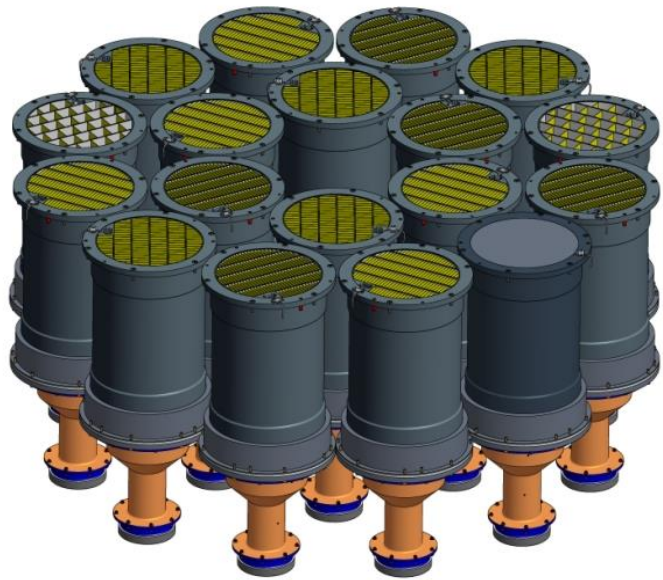

Figure 2: The structure of the HE main detector assembly

Each detector unit is a cylindrical NaI(Tl)/CsI(Na) phoswich scintillation detector with a diameter of 19 cm. The thickness of the main detector - NaI - is 3.5 mm, and of the shielding CsI is 40 mm. A 5-inch PMT is used to collect the fluorescence from both NaI and CsI (see Figure 2).

Because the sensitivity of the telescope highly depends on its background level, shielding design has to be introduced to reduce the background of either the diffuse ones or the high energy particles. Both the passive and active shielding techniques are applied to Insight-HXMT/HE. Apart from CsI(Na), HE is enveloped with plastic scintillator (veto) in order to reject the charged particle events. The veto consists of 18 plastic scintillators: six on the top and twelve surround the HE main detector assembly. Once the charged particle event triggers the scintillator plates, it will be rejected if there is a contemporary signal recorded in any of the 18 main detectors.

### 4.2 ME

ME contains three detector boxes, each with three units. One unit is subdivided into six modules, each ends up with 32 Si-PIN detector pixels. The modules work independently from each other and each is read out by one VA32TA2. Such a modularized design improves the overall reliability and makes the detectors easier to be installed. In short, ME has 1728 Si-Pin detector pixels which sum up to a total detection area of 952 $cm^2$. Figure 3 shows the schematics of one detector box and the detector array of the ME qualification model.

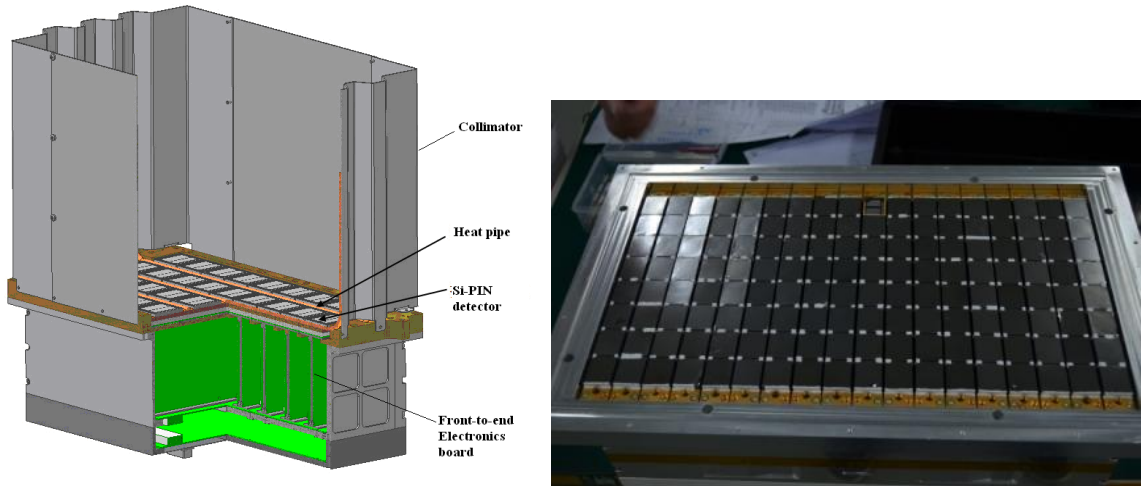

Figure 3: Schematics of one detector box (left) and the detector array of the ME qualification model (right).

### 4.3 LE

LE covers the energy band of 1.0-15 keV. It uses SCD as the detector so as to achieve a high time resolution. LE contains three detector boxes with a sun buffer each. The buffer could also be used as the radiator to cool the detectors. One detector box contains two modules and each module contains 16 SCD chips (CCD236). Every four CCD236 detectors share one collimator (see Figure 4). The total detection area of each module is 64 $cm^2$.

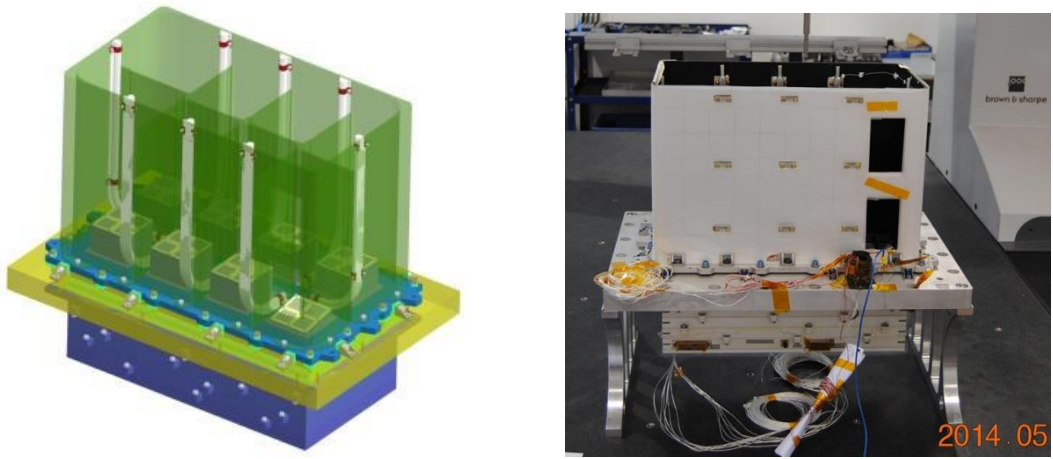

Figure 4: The sketch map of the LE detector module (left) and its qualification model.

# 5. MAIN CHARACTIRISTICS

The key parameters of the Insight-HXMT are listed in Table 1. A comparison of the sensitivity with other hard X-ray missions, and the effective area of Insight-HXMT are shown in Figure 5.

Table 1: The major characteristics of the Insight-HXMT

| Detectors | LE: SCD, 384 cm$^2$<br>ME : Si-PIN, 952 cm$^2$<br>HE : NaI/CsI, 5000 cm$^2$ |
|---|---|
| Energy Range | LE: 1-15 keV<br>ME: 5-30 keV<br>HE: 20-250 keV |
| Time Resolution | HE: 25 μs<br>ME: 280 μs<br>LE: 1 ms |
| Energy Resolution | LE: 2.5% @ 6 keV<br>ME: 14% @ 20 keV<br>HE: 19% @ 60 keV |
| Field of View of one module | LE: 6°×1.6°, 6°×4°, 60°×3°,blind<br>ME: 4°×1°, 4°×4°,blind<br>HE: 5.7°×1.1°, 5.7°×5.7°,blind |
| Angular Resolution (20σ source) | < 5' |
| Source Location (20σ source) | <1' |
| Sensitivity (3σ, in 10$^5$s) | 0.5 mCrab (only statistical uncertainties included) |
| Orbit | Altitude: ~550 km |
| Attitude | Inclination: ~43°<br>Three-axis stabilized |
| | Control precision: 0.1° |
| | Measurement accuracy: 0.01° |
| Data Rate | LE: 3 Mbps<br>ME: 3 Mbps<br>HE: 300 kbps |
| Payload Mass | ~1000 kg |
| Nominal Mission Lifetime | 4 years |
| Working Mode | Scan, pointing, GRB |

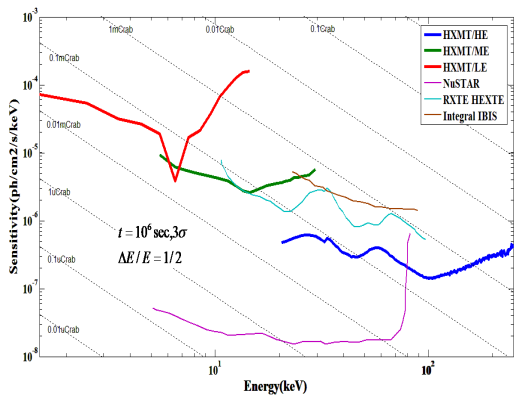
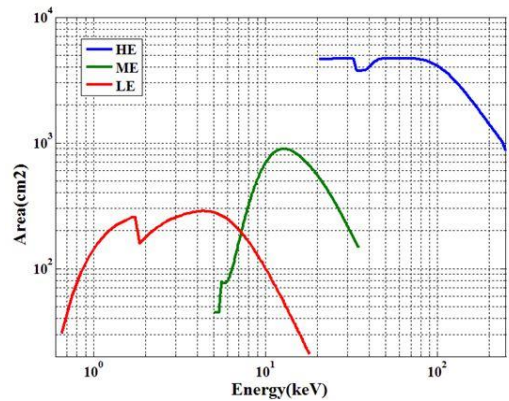

Figure 5: Left: Simulated sensitivities of the three telescopes of Insight-HXMT, based on pre-flight estimated background levels and without systematic uncertainties included. Also shown are the sensitivities of NuSTAR、RXTE/HEXTE and INTEGRAL/IBIS. The sensitivities of NuSTAR, INTEGRAL/IBIS and RXTE/HEXTE are reprinted from Koglin et al. (2005)[5].Right: The effective area of Insight-HXMT.

## 6. GROUND CALIBRATION

Reliable and accurate calibrations are the basis for a space telescope to achieve its scientific goals. For Insight-HXMT, the performance of the detector depends strongly on the working conditions. As exampled in Figure 6, both the energy resolution and the signal amplitude recorded by the LE detector are sensitive to the operating temperatures. So do the HE and ME detectors. The target of the calibration is to measure/estimate the response matrices of the telescope at different working conditions.

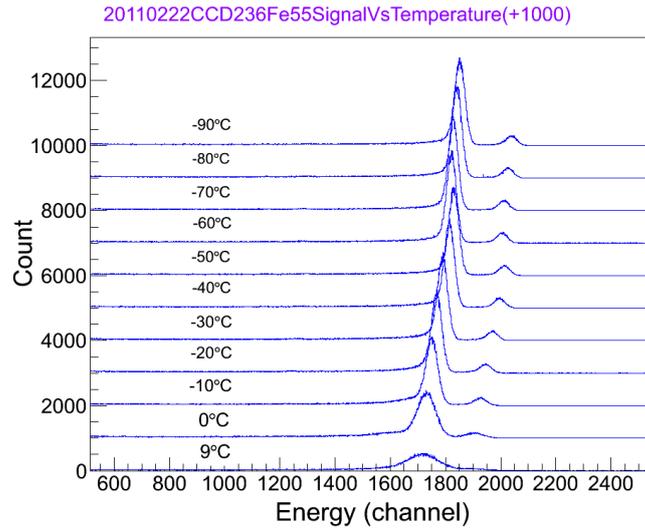

Figure 6: The $^{55}$Fe spectra measured by the SCD detectors at different temperatures

There are two calibration facilities built specially for the Insight-HXMT[6,7]. The facilities for HE works in 15-100 keV, for ME and LE works in 0.8-30 keV. For both the mono-energy X-ray beams are produced from the double crystal monochromators, with an intrinsic energy dispersion of roughly 0.1%-1%. Figure 7 shows the two facilities: the one for HE works in the air and under the room temperature, while the one for ME/LE is rather complex since supplementary devices like the vacuum chamber and cooling system have to be constructed to satisfy the working conditions of ME/LE.

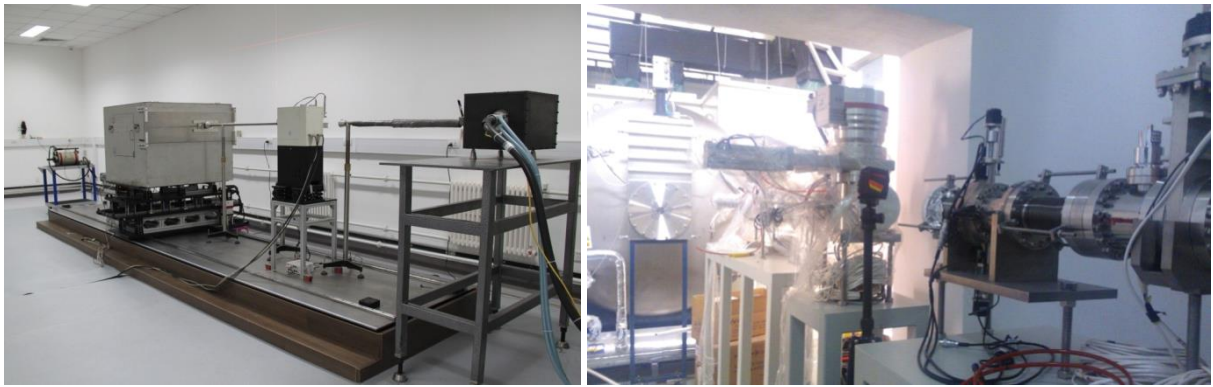

Figure 7: The calibration facilities for HE (left), and for ME/LE (right).

In addition to the above two facilities, there are two other vacuum chambers for the calibrations of ME and LE respectively. In the chambers, we use X-ray tubes to illuminate metal targets and hence produce florescent lines of iron, copper, molybdenum, and tin. These florescent lines are used to calibrate the variation of the energy response matrices of the detectors at the operating temperatures, in supplement to the monochromatic X-ray beams.

## 7. Insight-HXMT GROUND SEGMENT

### 7.1 Overview of Insight-HXMT ground segment

The Insight-HXMT ground segment (HGS) consists of two main parts. One is the mission operation ground segment (MOGS), which is taken care by the National Space Science Center (NSSC) of the Chinese Academy of Sciences (CAS). The functions of MOGS are about the operation commands, data reception, data pre-processing and data archive after the mission is finished. The science ground segment (SGS) is settled at the Institute of High Energy Physics (IHEP) of CAS, and covers four scientific centers to handle: user (HSUC), support (HSSC), operation (HSOC) and data (HSDC). See Figure 8 for details of the HGS, and in what follows the functions of SGS that are highly related to the scientific researches of Insight-HXMT are addressed in details.

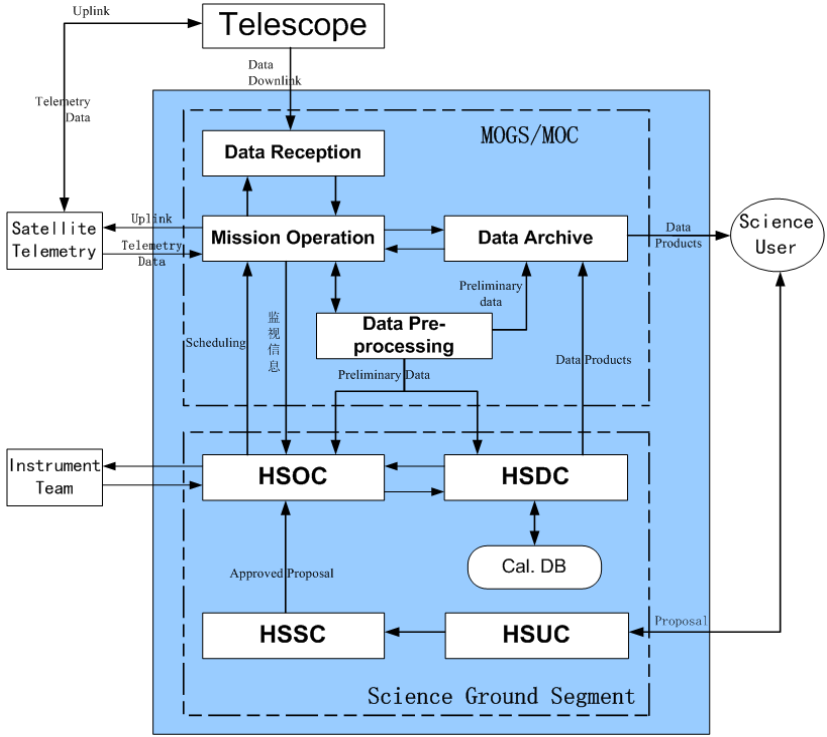

Figure 8: Overview of the Insight-HXMT HGS

### 7.2 Data archive

All the Insight-HXMT scientific data that will be released to the public are recorded in the standard FITS format. The standard scientific data products have three levels: level 0 is for the primary observational data, level 1 for data released to the public and level 2 for the scientific results obtained from the data. The HE data are classified into event, HE status, HE temperature, HE high voltage, count rate, dead time and data from the particle monitor. ME and LE data are made of event, status of the telescope, temperature, circuit parameters and the count rate. For more details see http://hsuc.ihep.ac.cn/FAQ/youknow.jsp and http://hsuc.ihep.ac.cn/web.

### 7.3 Analysis software

The Insight-HXMT Data Analysis Software (HXMTDAS) is developed in the software environment of HEAsoft, and is specified for analyzing the Insight-HXMT data from pointing observations. As shown in Figure 9, the input of HXMTDAS is the level 1 Insight-HXMT data products and outputs are the cleaned and calibrated event files and high-level scientific products of e.g. light curves and energy spectra. HXMTDAS is run in three steps: firstly to calibrate the event; then screen the event files and figure out the good time interval with standard event selection criteria; finally to extract the scientific data products ready for timing and spectral analysis with tools available in HEAsoft. The HXMTDAS and its guide for installation will be available at http://www.hxmt.org/index.php/enhome/analysis.

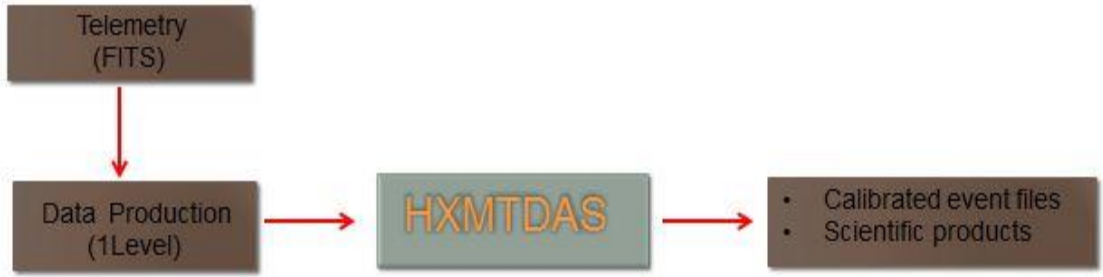

Figure 9: The input/output for HXMTDAS

### 7.4 AO-1 and core science team

During 2016 August and September, the first round of the Announcement of Opportunity (AO) was released to the national astronomers for calling submissions of the Insight-HXMT core science proposals. After been evaluated in both technical and scientific aspects, 90 proposals were approved by the Insight-HXMT scientific committee. These proposals request observations of over than 300 targets, which sum up to exposures of 331 days for normal observations and 1140 days for TOO observations. As shown in Figure 10, half of the requested exposure will be spent on the high cadence and high statistics observations of BH and NS XRBs. The Insight-HXMT core science team was formed with its members mainly from Insight-HXMT project, and the PIs/Co-Is of these core science proposals. The core science team is divided into eight groups to address the core sciences in the following aspects:

  Group 1: Accretion X-ray binaries
  Group 2: Galactic survey and diffuse emission
  Group 3: Multi-wavelength observations
  Group 4: Calibration and background model
  Group 5: Pulsar navigation
  Group 6: Extragalactic sources
  Group 7: Gamma-ray bursts and gravitational wave EM counterparts
  Group 8: Non-accreting pulsars

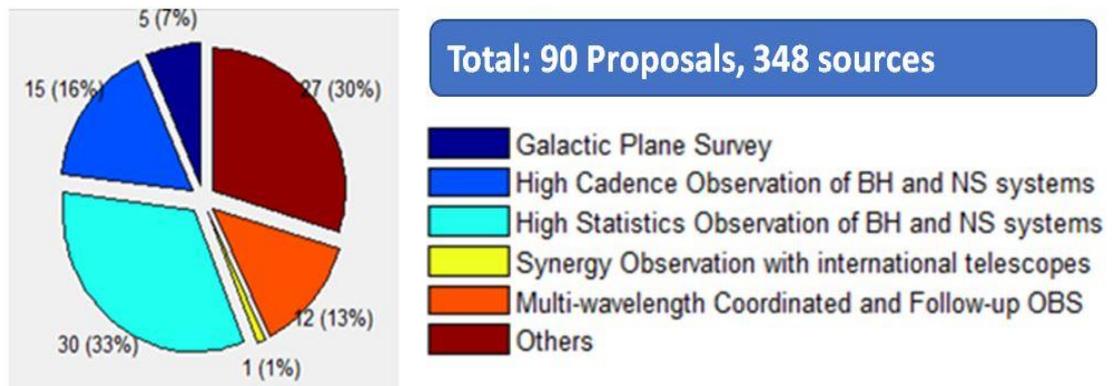

Figure 10: The research fields that are covered with the core science proposals approved in AO-1

## 7.5 Data policy

There are two major research programs for Insight-HMXT in-orbit observation, data analysis and associated research. 1) Insight-HXMT Core Science Program. The observational data for core science program will be shared within the core science team. 2) Insight-HXMT Guest Science Program. Insight-HXMT Guest Science Program is formulated by guest scientists from both domestic and foreign research group, where users will be able to propose observations not already included in the Core Science Program. The approved proposals will be scheduled by Insight-HXMT SGS, while the observations will be exclusive for guest science group during the data protection phase, including the data analysis and scientific paper publication.

## 7.6 Observational modes

Insight-HXMT has three observational modes: pointing, scan and GRB. The pointing observation will have duration from one orbit (96 mins) up to 20 days, dedicated to spectral and timing researches. The Galactic plan will be surveyed with the small area scan mode. The entire Galactic plan is divided into 22 patches, each with a size of 20 deg×20 deg. The scan of one patch (see Figure 11) takes 2 hours ~ 5 days depending on combinations of the different scan parameters. GRB mode was designed and implemented for HXMT/HE before launch. In this mode, the high voltage of the photo-multiplier tube (PMT) is reduced, so that the measured energy range of CsI can be increased to 0.2-3 MeV. This makes HE a unique high-energy gamma-ray telescope because of a FOV of almost the entire sky, a large collection area (>1000 cm$^2$) and microsecond time resolution.

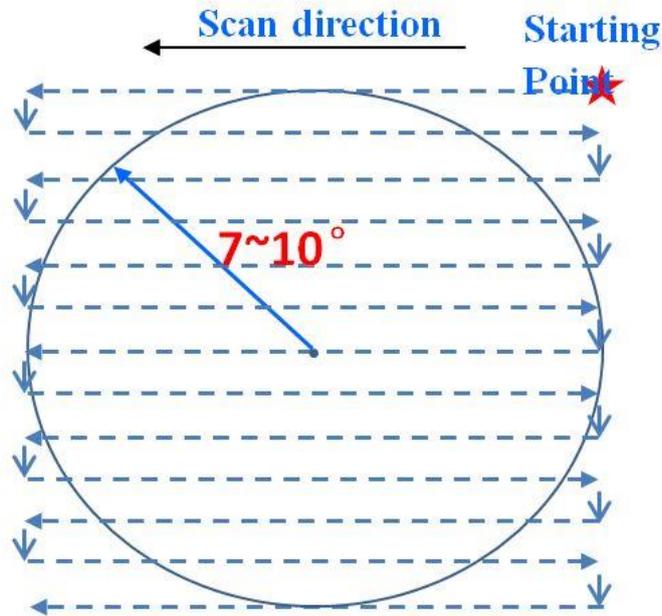

Figure 11: Demonstration of how a small sky area is scanned with Insight-HXMT.

## 7.7 Observational plan and constraints

Insight-HXMT SGS prepares the observational plan in term of the long, medium and short periods according to the corresponding science programs and the prediction of the orbit. These plans are made for observational periods of 1 year, 4 weeks and 2 days, with accuracy of up to 2 weeks, 2 days and 1 second, respectively. Making an observational plan is subject to the following strict constraints. 1 thermal control: the solar avoidance angle > 70 °, and X-Z plane of the payload < 10 ° (see Figure 12); 2 earth occultation: lasts about 30 minutes for almost every orbit; 3 south atlantic anomaly (SAA): roughly 8~9 orbits per day, with each lasting for about 15 minutes; 4 moon avoidance angle: > 6 °.

The Insight-HXMT observational plans are available at http://www.hxmt.org/index.php/enhome/schedule. Please note that the Insight-HXMT short term plans are subject to frequent modifications due to the capability of Insight-HXMT's fast response to TOO observations.

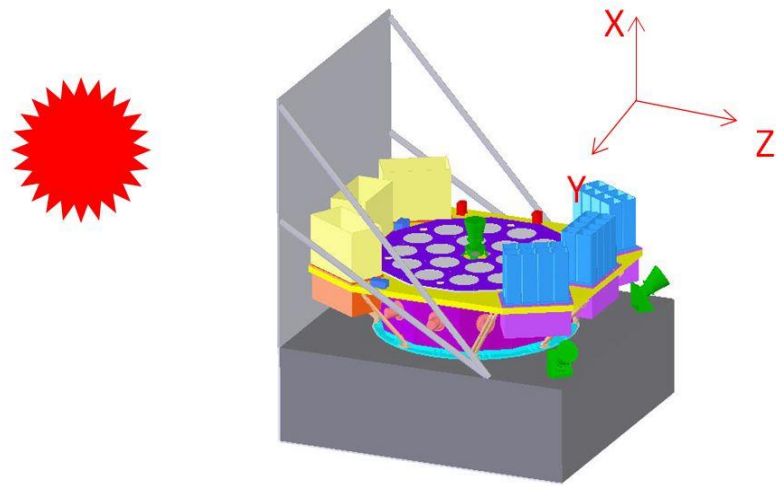

Figure 12: The consideration of the thermal control of Insight-HXMT

### 7.8 TOO strategy and fast response of Insight-HXMT

Once Insight-HXMT receives a TOO trigger from either the Insight-HXMT small area survey or from the astronomy society, the SGS will active a procedure for fast response and arrange accordingly a TOO observation. The entire flow chart of Insight-HXMT TOO strategy is shown in Figure 13, from which one sees that Insight-HXMT can response to a TOO trigger and perform a TOO observation within a time period of as short as 5 hours.

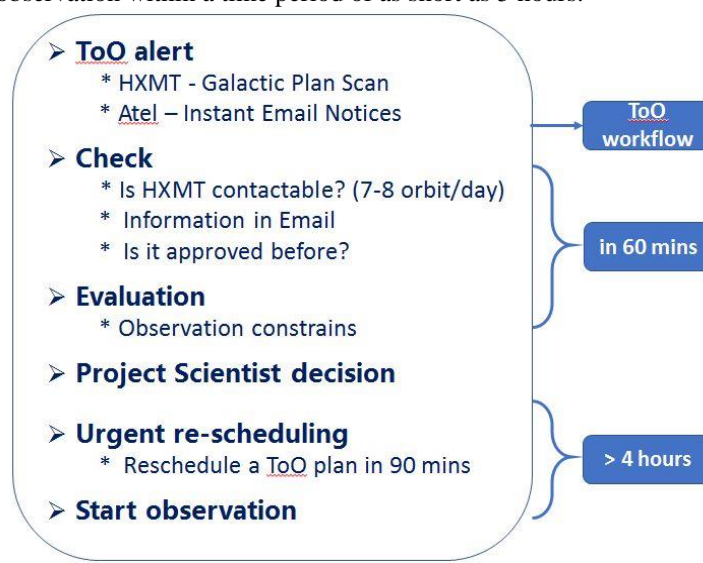

Figure 13: Insight-HXMT TOO response flow chart

## 8. IN-ORBIT PERFORMANCE

After Insight-HXMT was launched in JiuQuan at 11 am local time of June 15$^{th}$ 2017, six hundred seconds later the launcher delivered successfully the satellite to the preset orbit (Figure 14). The platform was firstly tested in the next nine days, and so did the performances of the payload and the ground segment during the following three months. Shown in Figures 15-17 are a few examples for in-orbit performances of HE, ME and LE: the dead time measurement of HE (Figure 15), the comparisons of the spectra measured in-orbit and on ground for pixels illuminated by the calibration radioactive sources (Figure16 For ME and Figure 17 for LE). Also the SAA region is well mapped by the particle detectors during 2017 June 15 – July 11 (see Figure 18). The performance verification phase shows that the entire satellite works smoothly and healthily, and Insight-HXMT started to be in service for scientific observations at the end of 2017.

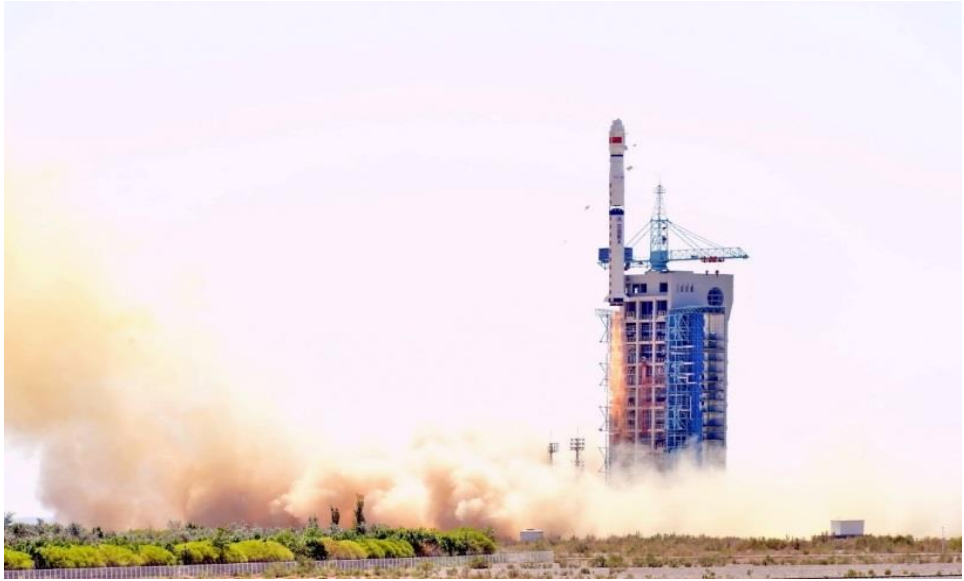

Figure14: Launch of Insight-HXMT at Jiuquan, China

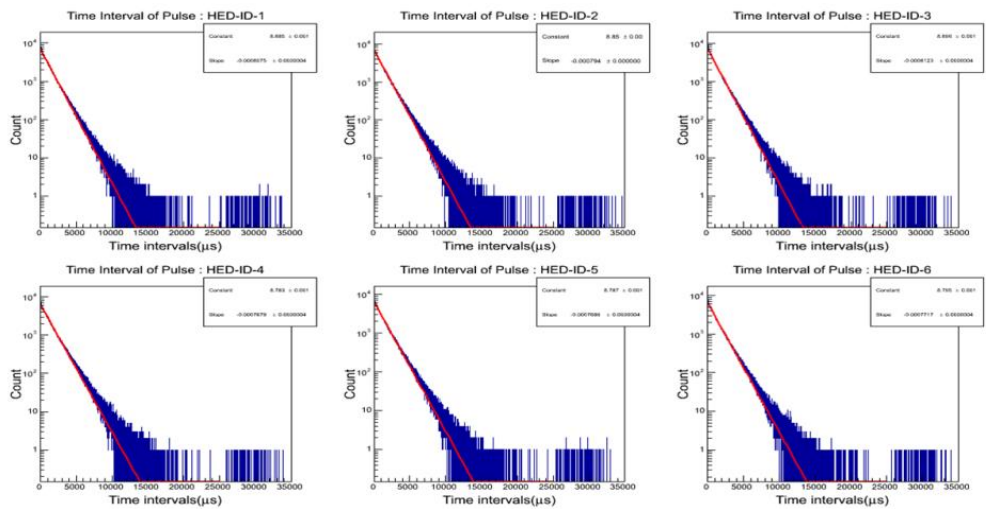

Figure 15: Measurement of the dead time for HE in orbit

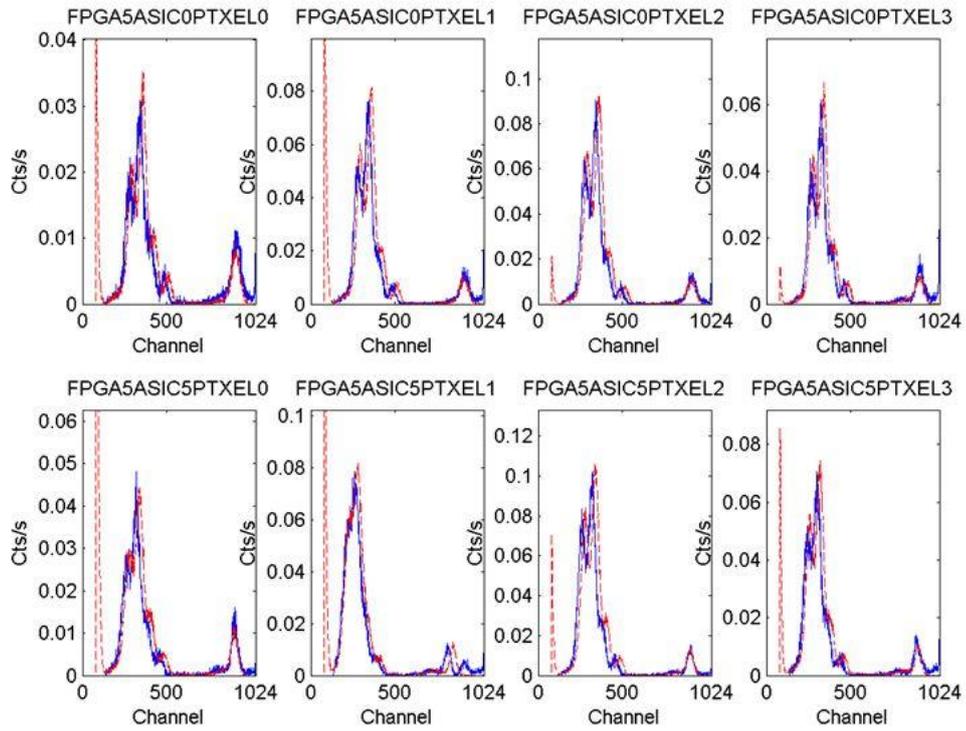

Figure 16: Comparisons of the spectra measured in-orbit and on ground for ME pixels illuminated by the calibration radioactive sources

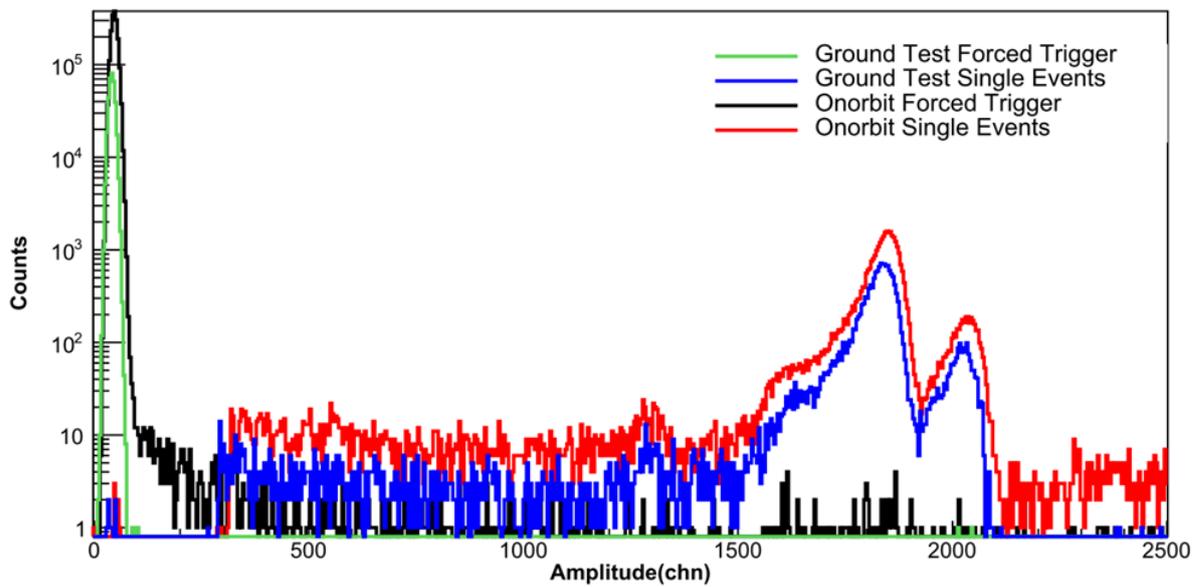

Figure 17: Comparisons of the spectra measured in-orbit and on ground for LE pixels illuminated by the calibration radioactive sources

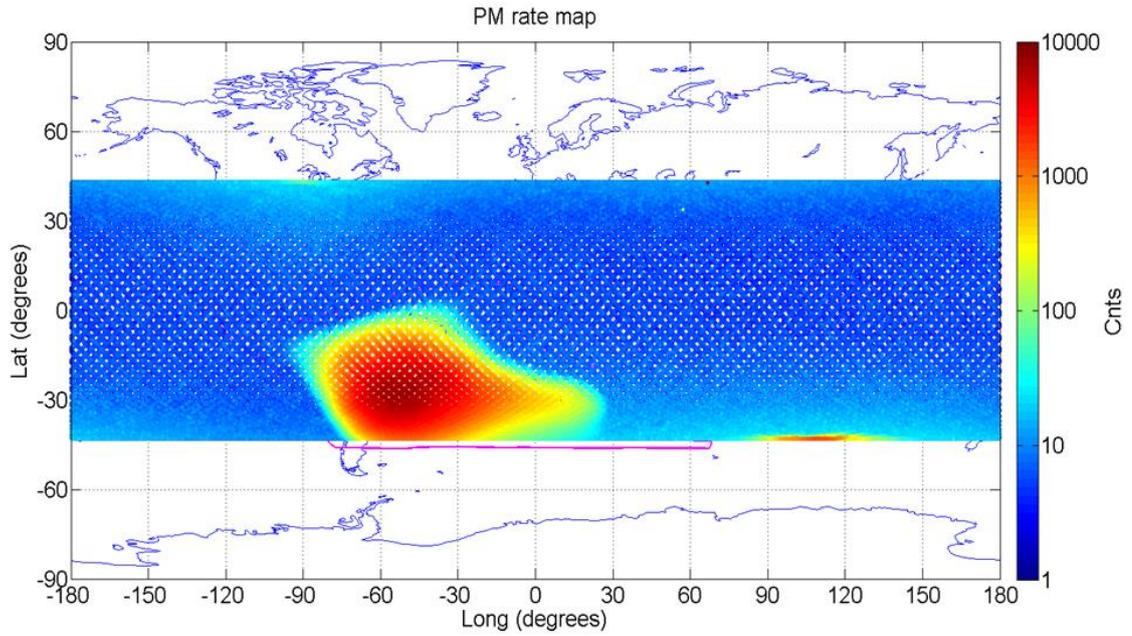

Figure 18: SAA region mapped by the Insight-HXMT particle detectors

## 9. IN-ORBIT CALIBRATION

The Insight-HXMT in-orbit calibrations include mainly the aspects of: timing system, point spread function (PSF), energy response, effective area, and calibration in GRB mode.

Calibration of the timing system needs pulse-profile observed for the isolated pulsars. Although the stable pulsar PSR 1509-58 was observed two weeks after the launch, it is too weak compared to Crab to have sufficient statistics for timing calibration. The major calibrations were carried out when Crab became visible since August 27 2017. As shown in Figure 19, the pulse profiles are derived for HE, ME and LE, which result in the time accuracy of 22 μs, 51 μs, and 21 μs for HE, ME and LE, respectively. The systematic shift in pulse profile of LE is due to the special read-out system intrinsic to LE (Figure 20), which is well understood and hence the phase shift can be properly corrected.

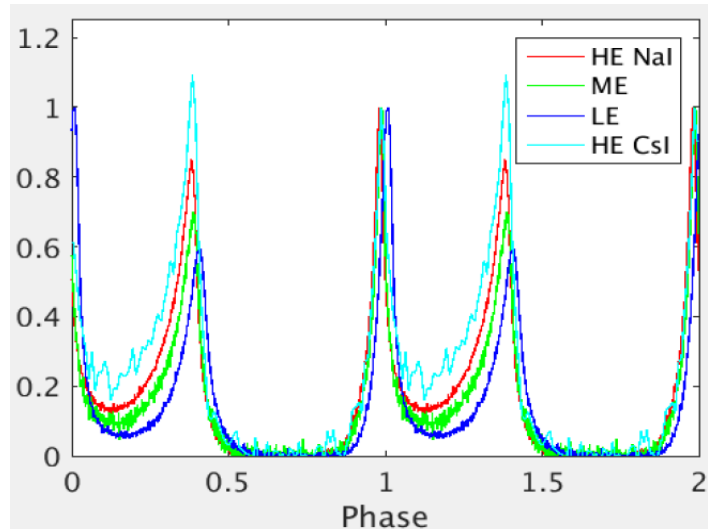

Figure 19: The pulse profile of Crab as observed by HE, ME and LE

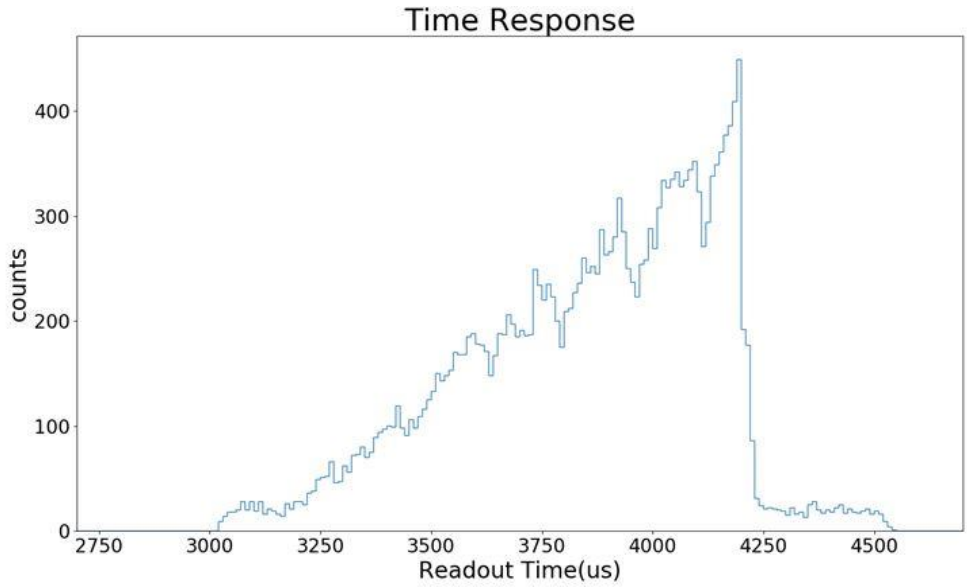

Figure 20: LE time response

Crab was observed in a scan mode for PSF calibration of Insight-HXMT. In the scan mode, Crab shows up in the light curve as a sequence of triangles (Figure 21). The PSF was revised in order to fit well the source signal and reduce the residuals after PSF fitting (Figure 22).

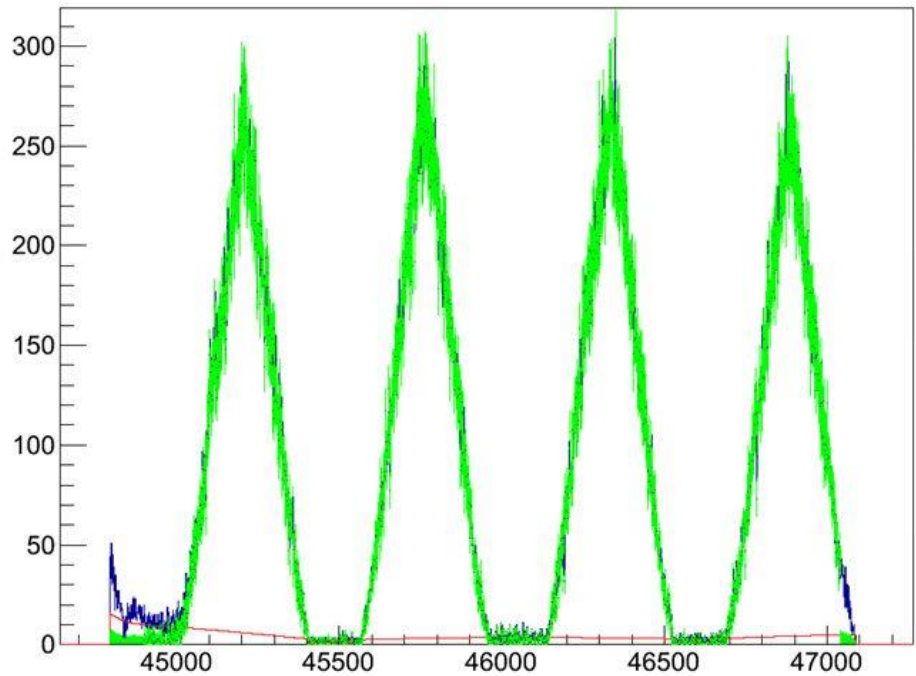

Figure 21: Crab light curve observed by LE in a scan mode. The red line shows fit to the background.

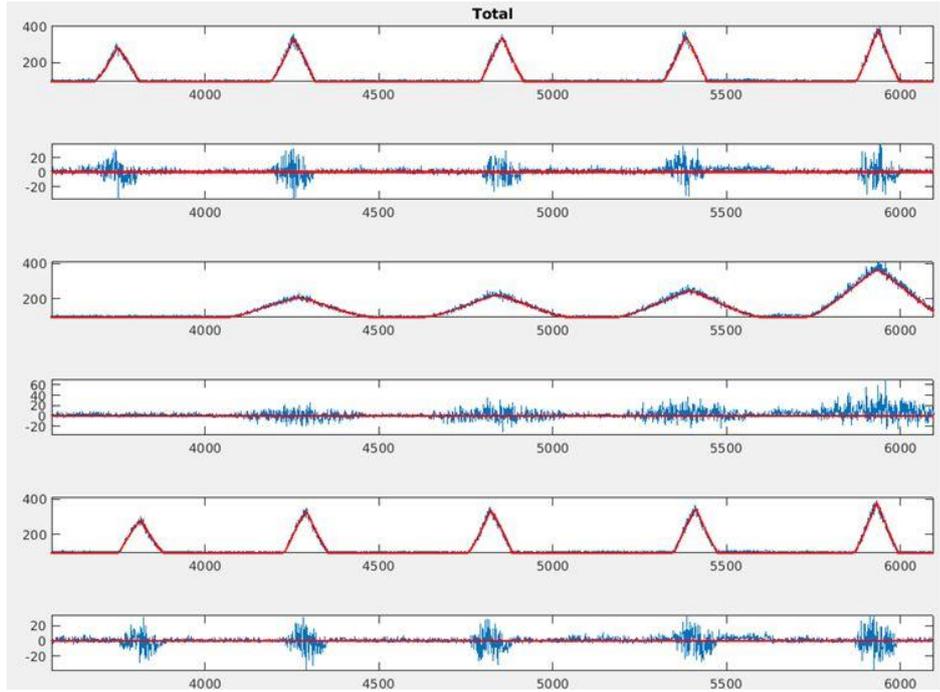

Figure 22: The PSF fit of the Crab light curve observed by LE in a scan mode. Each pair of the figure denotes the light curve recorded by one LE detector box and the residual after PSF fitting to the source signal.

The calibration of the energy response consists of the energy-channel (EC) relation and the energy resolution. The energy resolution can be investigated via analyzing the line emission for pixels illuminated by the in-orbit radioactive sources and the results show that the energy resolutions are consistent with those values listed in Table 1. The EC calibration needs emission lines in spectrum and the calibrations are carried out in combination with the EC relation measured on the ground with the calibration facilities. For LE, the observations of Cas A provide six lines which are used to adjust EC (Figure 23). For ME there is only silver line visible in the background spectrum. Therefore, the overall property of EC relation for ME is firstly evaluated from over than 100 pixels that are illuminated by the in-orbit radioactive source Am 241 (Figure 24), and then compared to the EC measured on the ground. The EC relation for majority of the ME pixels are adjusted based on their detections of the silver line. For HE four lines from background spectrum are taken for EC calibration (Figure 25).

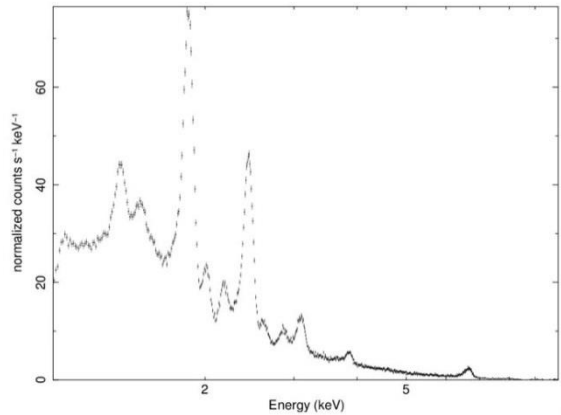

Figure 23: The Cas A spectrum observed by LE

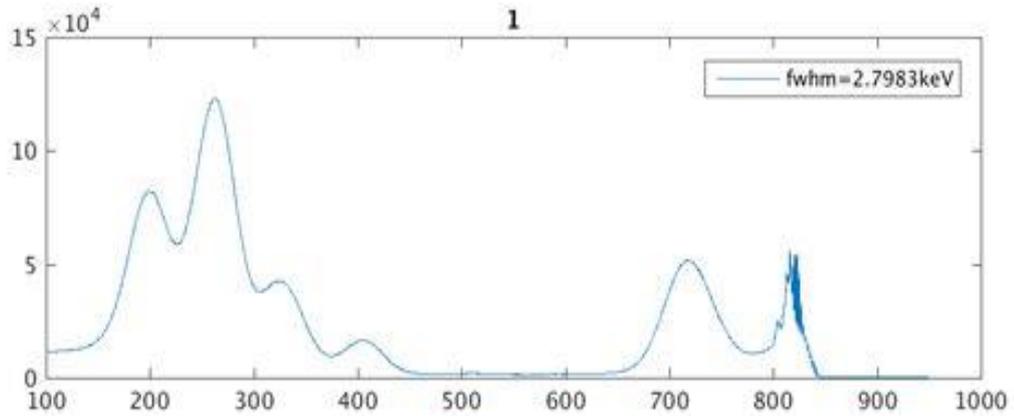

Figure 24: ME spectrum for pixels illuminated by the in-orbit radioactive source Am241

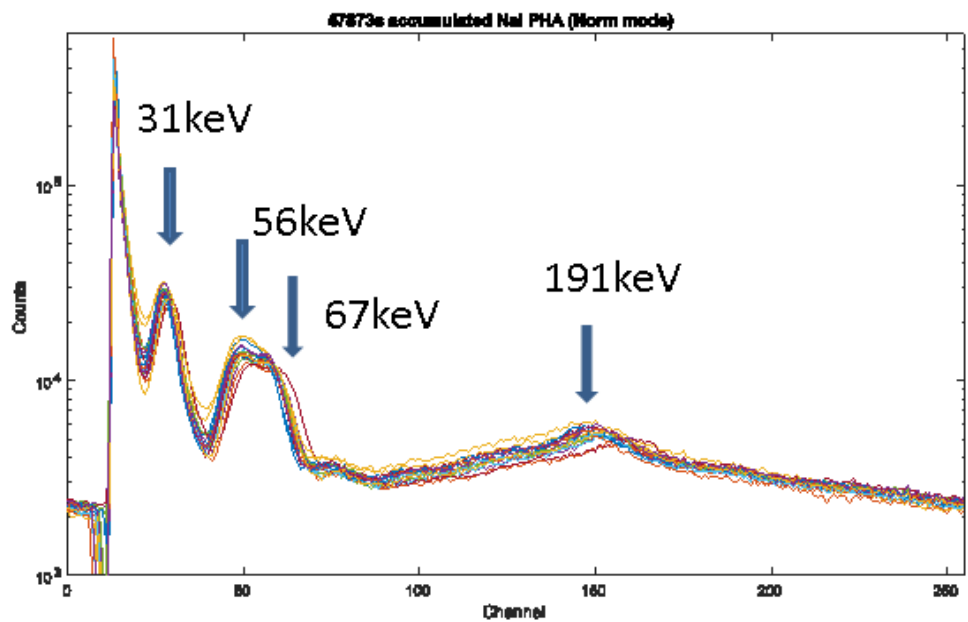

Figure 25: The background spectrum with four lines which are used to verify/adjust EC of HE

The strategy of ARF calibration is to take the known spectral shape of the source and to disentangle the uncertainty contributed by background modelling. Since for collimated telescope the background modelling is usually a challenging and heavy task, which is currently ongoing in parallel to the ARF calibration for Insight-HXMT, we take the pulse-off emission of Crab as the background and calibrate the ARF with Crab pulse-on observation. The Crab pulse-on spectrum is obtained mainly from RXTE observations of Crab and is cross-checked for consistency with that observed by Bepposax and Nustar. In the spectral fitting procedure, the Crab pulse-on spectral parameters are fixed, and the ARF calibration is performed via introducing an additional function to represent the residual of the spectral fitting. With the known Crab pulse-on spectrum the joint fit for ARF calibration is shown in Figure 26.

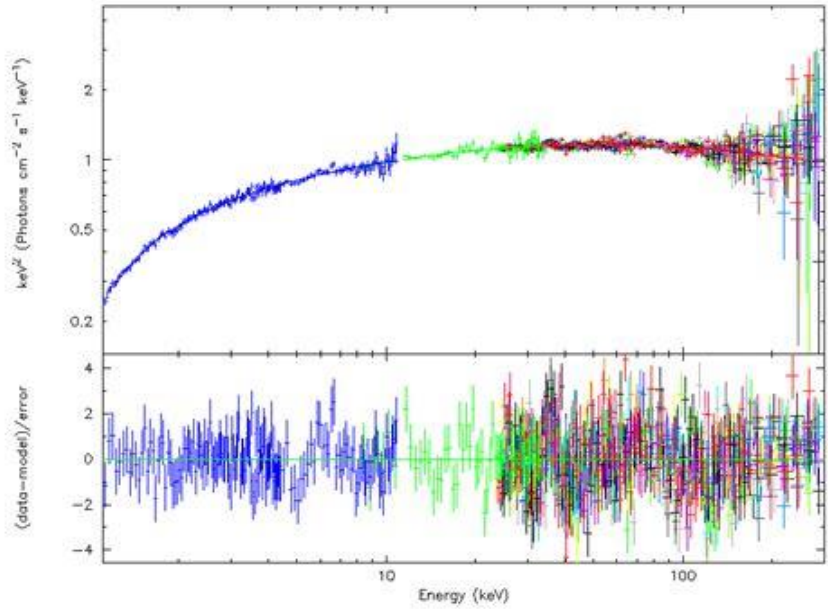

Figure 26: The joint fit for ARF calibration, with the fixed pulse-on spectrum of Crab.

The in-orbit calibration of CsI in GRB mode turns out to be a tough task. Some of the incident GRB photons will be blocked by the platform structures and the CsI detectors can map photon distribution at a given incident direction of GRB event. The calibrations are carried out mainly based on the GEANT4 simulation and the Crab observation. The simulations show that, by regarding the structure of the platform as a coded mask, GRB can be localized to an accuracy of a few degrees (Figure 27).

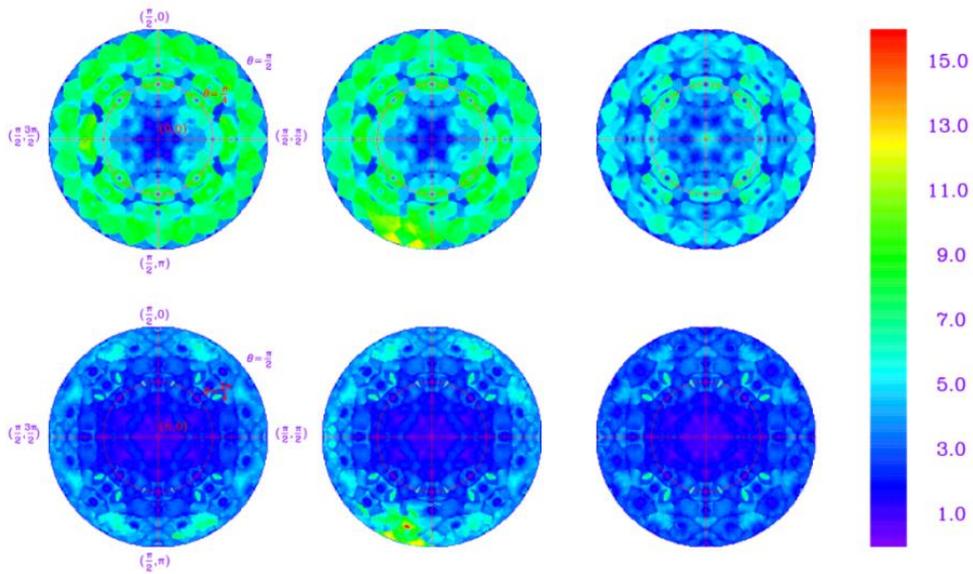

Figure 27: Localization capability of HE/CsI in GRB observation. Each map shows the photon distribution recorded by CsI for a specific GRB incident direction.

## 10. IN-ORBIT BACKGROUND

For a collimated telescope the background is usually measured by taking an on/off observational mode. This is however relatively hard for Insight-HXMT because the particle-induced background changes significantly with time, and also due to the fuel consumption and the complexity of the sky region at soft X-rays. Additionally, an on/off mode will lose roughly half of the observational time. We therefore adopted a design to measure the contemporary background: some modules are set blind for estimating the instrumental background. Accordingly, the background models are built by taking the correlations of the count rate between the blind and other detectors. Observations of a series of blank sky regions are the necessary input for building and testing the background model. Figure 28 demonstrates how the background of a blank sky observation is recovered from the background model built for one HE module.

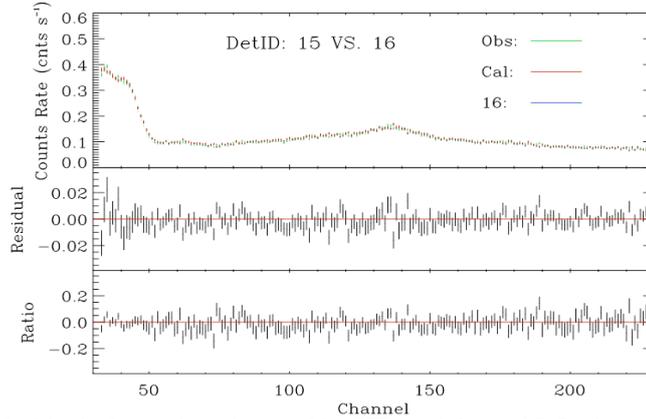

Figure 28: The HE spectrum of a blank sky region observed by HE module no.15. In the upper panel, the green points are the observational data, the blue points are the contemporary background measured by the blind module no.16, and the red points are background estimated based on the background model. The lower two panels show residual of the spectral fit and ratio of the count rate between no.15 and the estimated background from model.

## 11. OBSERVATIONS

Till 2018 April, Insight-HXMT has been in service for about 300 days. A variety of observations have been carried out by Insight-HXMT, which include over than 500 pointings and roughly 300 times for survey of the small sky regions. A summary is given in Figure 29 in details and the sky explored so far by Insight-HXMT is shown in Figure 30.

| # | Mode | Type | Source Name | pointings | exposure |
|---|---|---|---|---|---|
| 1 | Point | Pulsar remnants | Cas A | 9 | 530 |
| 2 | | | Crab | 85 | 1520 |
| 3 | | Pulsar | PSR B0540-69 | 7 | 250 |
| 4 | | | PSR B1509-58 | 12 | 310 |
| 5 | | | Cyg X-1 | 9 | 220 |
| 6 | | | Granat 1716-249 | 2 | 250 |
| 7 | | | GRS 1915+105 | 6 | 260 |
| 8 | | | GX 339-4 | 1 | 100 |
| 9 | | BH Binary | H 1743-322 | 15 | 180 |
| 10 | | | MAXI J1535-571 | 18 | 430 |
| 11 | | | MAXI J1543-564 | 1 | 80 |
| 12 | | | MAXI J1820+070 | 10 | 410 |
| 13 | | | Swift J1658.2-4242 | 23 | 430 |
| 14 | | | 1ES 1959+650 | 25 | 255 |
| 15 | | Extra-galactic | Perseus | 2 | 200 |
| 16 | | | M87 | 4 | 180 |
| 17 | | | Cosmos Field | 4 | 80 |
| 18 | Point | | Aql X-1 | 3 | 30 |
| 19 | | | Cen X-3 | 11 | 350 |
| 20 | | | Cir X-1 | 6 | 100 |
| 21 | | | Cyg X-2 | 12 | 320 |
| 22 | | | Cyg X-3 | 11 | 340 |
| 23 | | | GRO J1008-57 | 11 | 340 |
| 24 | | | GRO1750-27 | 1 | 15 |
| 25 | | | GS 1826-238 | 1 | 40 |
| 26 | | | GX 301-2 | 12 | 340 |
| 27 | | NS Binary | GX9+9 | 2 | 50 |
| 28 | | | GX 17+2 | 5 | 150 |
| 29 | | | Her X-1 | 6 | 250 |
| 30 | | | Sco X-1 | 3 | 150 |
| 31 | | | Vela X-1 | 1 | 120 |
| 32 | | | 2A 1822-371 | 1 | 30 |
| 33 | | | 4U 1728-34 | 3 | 70 |
| 34 | | | 4U 0115+63 | 11 | 150 |
| 35 | | | 4U1636-536 | 17 | 170 |
| 36 | | | PSR J2032+4127 | 4 | 40 |
| 37 | | TBD | Swift J0243.6+6124 | 97 | 1200 |
| 38 | | BlankSky | | 73 | 700 |
| 39 | SAS | Crab Area | | 9 | 550 |
| 40 | | Galactic Plane | 22 regions | 284 | 3200 |

Figure 29: Summary for Insight-HXMT observations carried out till 2018 April. Here the exposure is in units of ks. The sources in red stand for the TOO observations and in black for the normal pointing observations.

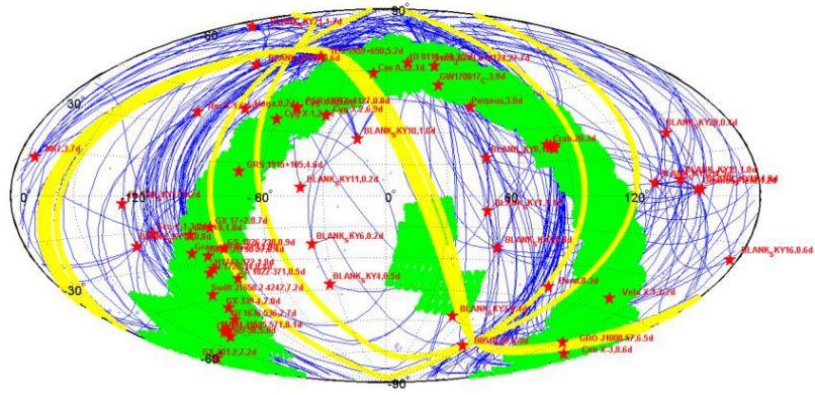

**Red stars:** point observation
**Green regions:** small area scan

Figure 30: The all sky coverage of Insight-HXMT till 2018 April. Here the red stars are the pointing observations, the green belts are the small area scan, the yellow belts are the test of the all-sky survey mode, blue lines are tracks of slew between different observations.

## 12. PRELIMINARY RESULTS

Preliminary results obtained by Insight-HXMT cover a rather wide scope: the Galactic plane survey, the timing and spectral analyses of outburst BH XRB and BS XRB systems, the timing and spectroscopy of pulsars, bursts from low mass XRB, flaring AGN, observation of EM counterpart of the NS-NS merger event GW170817, hunting for the GRB at domain of hard X-rays to soft gamma-rays, and so on.

Here we give a few examples for those preliminary results derived so far by Insight-HXMT. As shown in Figure 31 is the Galactic center region mapped by applying the DDM imaging method to LE observation. Figure 32 shows the low frequency QPOs detected by HE, ME and LE in a newly discovered BH XRB transient MAXI J 1535-571. For the newly discovered NS XRB transient Swift J0243.6+6124, the pulse profile of the neutron star shows complicated patterns along with the outburst evolution (Figure 33). During the famous GW EM observational campaigns, Insight-HMXT was among the only four X/gamma-ray telescopes monitored the GW source throughout the trigger time (Figure 34)[8], and hence reported the most stringent constraints upon the emission at hard X-rays – soft gamma-rays (Figure 35) for GW170817 during time periods of the precursor, the GBM trigger, and the after glow[9]. More than 60 GRBs were detected by HXMT/HE and Figure 36 shows one example for the better sensitivity of HE with respect to SPI and GBM in hunting for the short/hard GRB event.

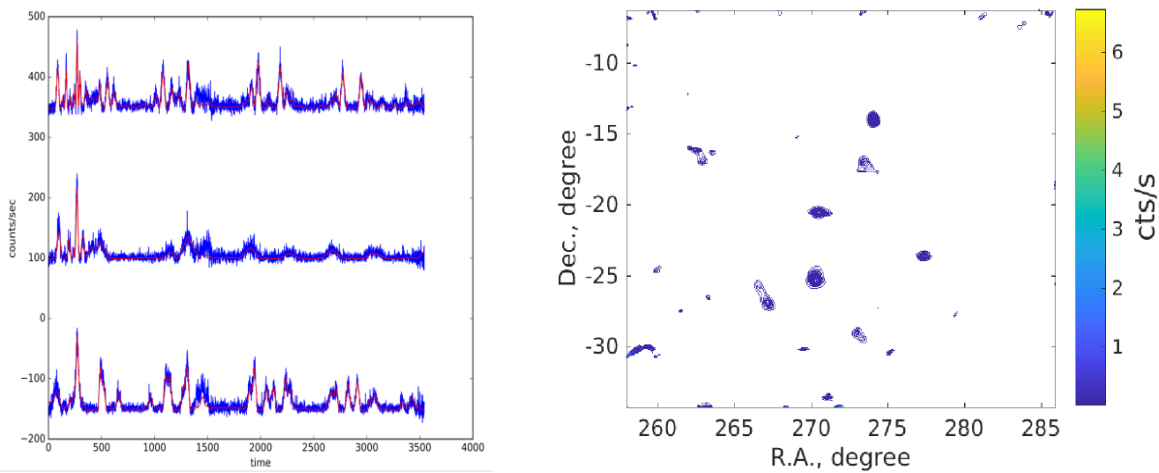

Figure 31: LE observation of the Galactic cernter region: left is the scan light curve and right is the sky map built with the direct demodulation method.

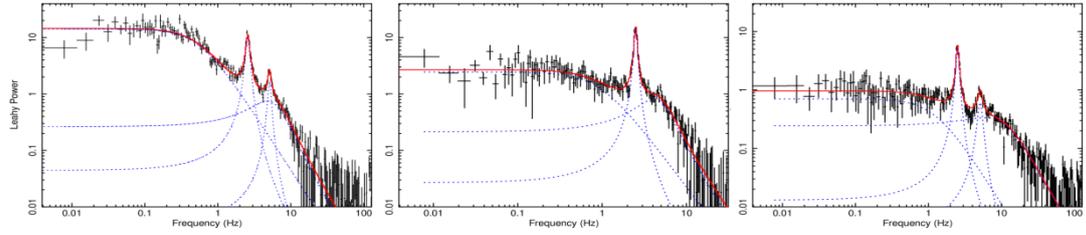

Figure 32: The low frequency QPOs detected by HE (right), ME (middle) and LE (left) under an exposure of 3ks, during the low/hard state of the outburst from the newly discovered BH candidate MAXI J1535-571.

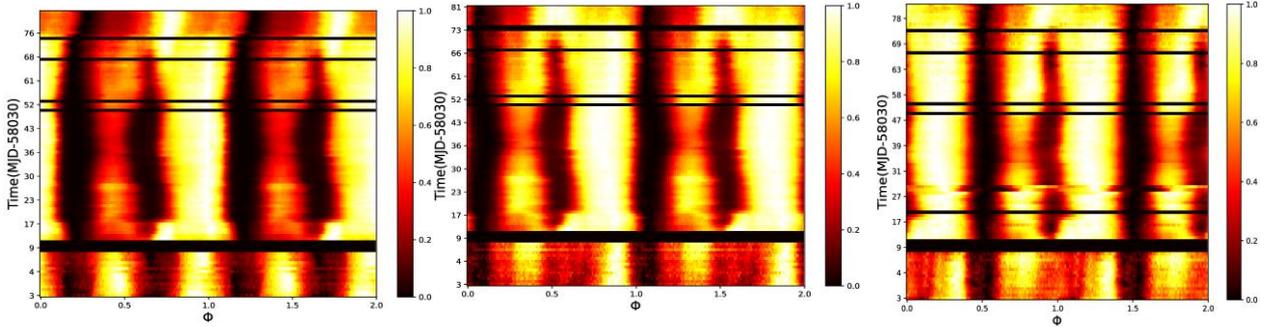

Figure 33: Pulse profile evolution of the NS star harbored within the newly discovered NS XRB system Swift J0243.6+614 during the outburst. From left to right are the pulse profiles recorded by HE, ME and LE, respectively.

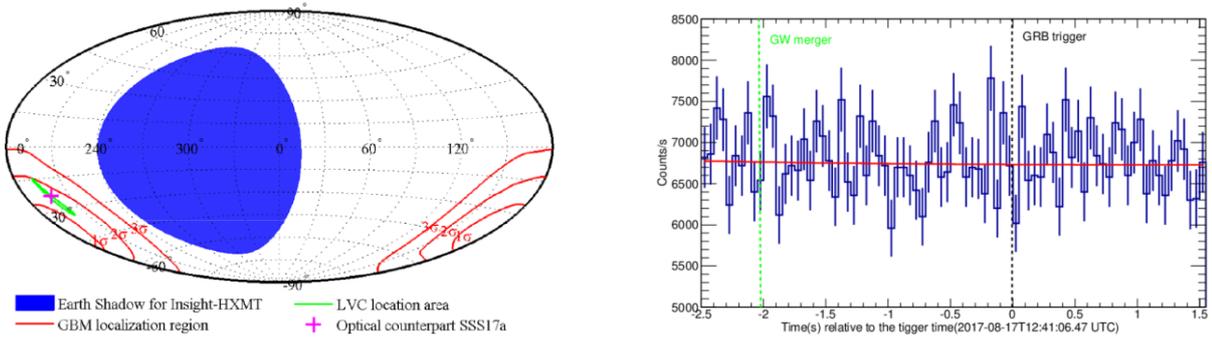

Figure 34: The sky region of GW170817 monitored by Insight-HXMT/CsI (left), and the light curve (right) covers both time perods around the GW merger and GRB trigger.

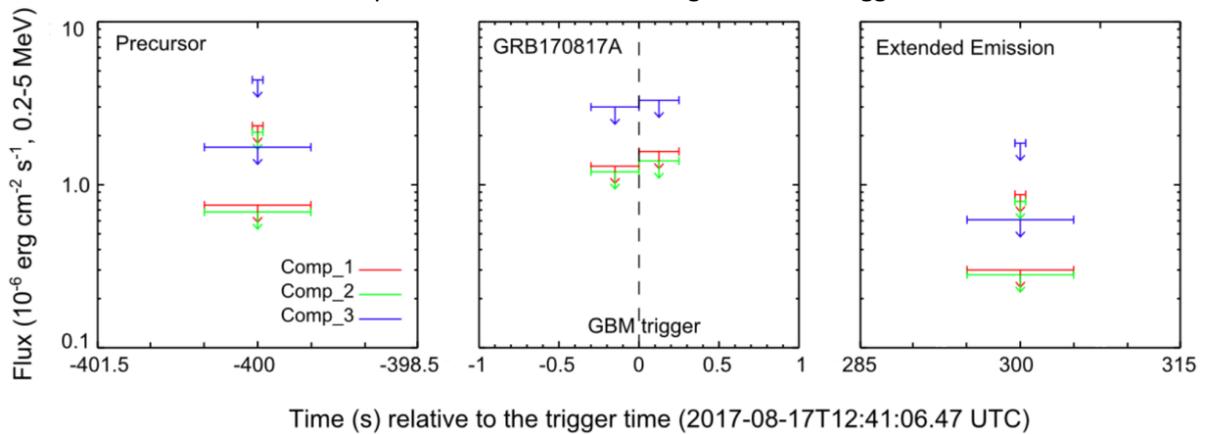

Figure 35: The Insight-HXMT detections of the flux limits of the EM counterpart of GW170817 during precursor, GRB trigger and for the extended emission [9].

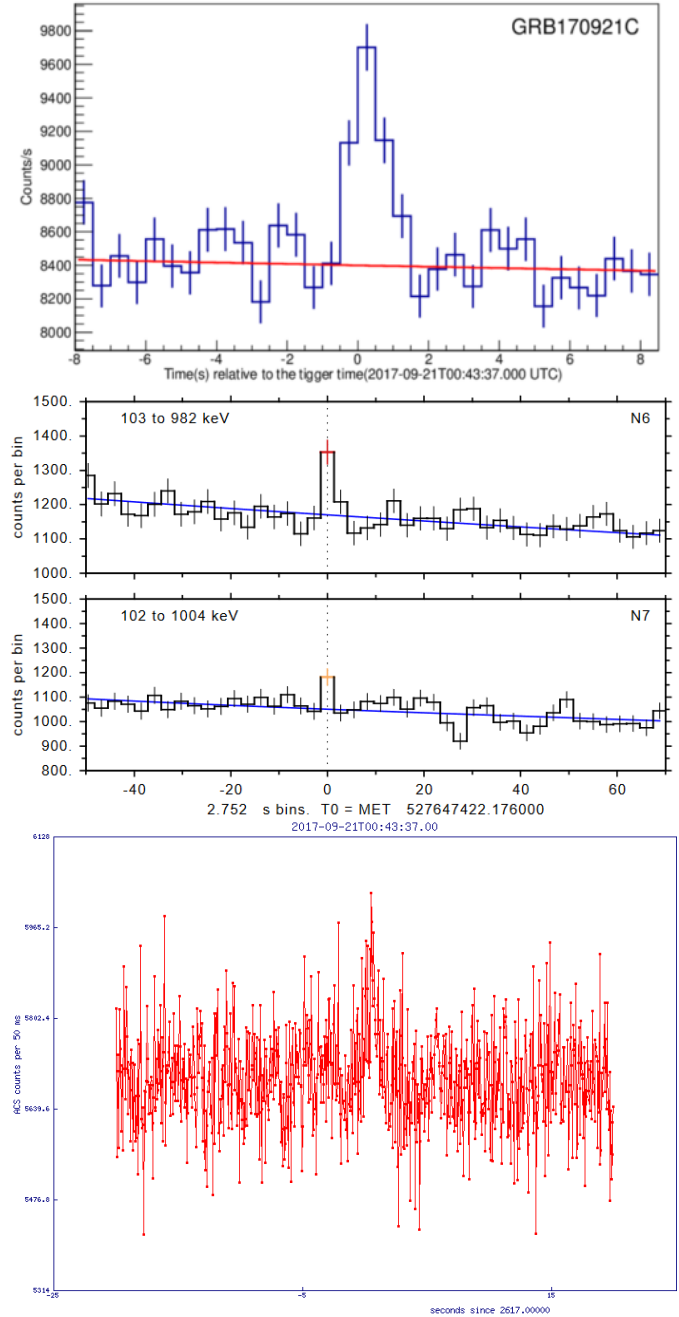

Figure 36: GRB 170921C detected by Insight-HXMT (upper, 12 σ), Fermi/GBM (middle, 8 σ) and SPI-ACS (lower, 4 σ).

## 13. SUMMARY

As China's 1st X-ray astronomy satellite, the detectors of Insight-HXMT cover energy bands of 1-15, 5-30, 20-250 keV and 200-5000 keV (all-sky monitor mode of GRB). Insight-HXMT finished its performance verification and calibration phases during a time period from June 15 to November 15 2017, and entered into normal observation phase by the end of 2017. So far it has carried out Galactic plane scan and monitoring, performed normal/TOO observations on many bright sources. More than 60 GRBs have been detected and the GW EM follow-up observations put stringent constraints at soft gamma-ray energies for the counterpart of GW170817. GRB mode is usually switched on in a band pass of 0.2 to 5 MeV when Insight-HXMT moves into the Earth shadow or HE is not used. Collaborations are very welcome and are foreseen in three ways: 1) partner institutions that had provided contributions to Insight-HXMT; 2) coordinate multi-wavelength observations of both in space and on ground; 3) fill the specific application form to become a core science team member of Insight-HXMT.

## ACKNOWLEDGEMENTS

This work made use of the data from the Insight-HXMT mission, a project funded by China National Space Administration (CNSA) and the Chinese Academy of Sciences (CAS). The authors thank support from the National Key Research and Development Program of China (2016YFA0400800), XTP project XDA 04060604, the Strategic Priority Research Programme 'The Emergence of Cosmological Structures' of the Chinese Academy of Sciences, Grant No.XDB09000000, and the Chinese NSFC 11473027, 11733009, 11373036 and 11133002.